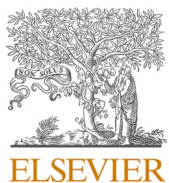

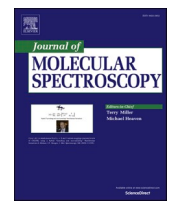

# Absorption cross-sections for the 5th and 6th vibrational overtones in a series of short chained alcohols using incoherent broadband cavity enhanced-absorption spectroscopy (IBBCEAS)

Callum E. Flowerday [a, 1], Nitish Bhardwaj [a, 1], Ryan Thalman [b], Matthew C. Asplund [a], Eric T. Sevy [a], Jaron C. Hansen [a, *]

[a] Department of Chemistry and Biochemistry, Brigham Young University, Provo, UT 84602, USA
[b] Department of Chemistry, Snow College, Richfield, UT 84701, USA

### A B S T R A C T

Absorption cross-sections for the 5th $(6 \leftarrow 0)$ and 6th $(7 \leftarrow 0)$ OH overtones for gas-phase methanol, ethanol, and isopropanol were measured using a slow flow cell and Incoherent Broadband Cavity-Enhanced Absorption Spectroscopy (IBBCEAS). Measurements were performed in two wavelength regions, 447–457 nm, and 508–518 nm, using two different instruments. The experimental results are consistent with previous computational predictions of the excitation energies for these transitions. Treating the OH stretch as a local mode allowed for calculation of the fundamental vibrational frequency ($\omega_e$), anharmonicity constant ($\omega_e x_e$), and the vertical dissociation energy (VDE) for each alcohol studied. The fundamental vibrational frequency is $3848 \pm 18$ cm$^{-1}$, $3807 \pm 55$ cm$^{-1}$, and $3813 \pm 63$ cm$^{-1}$ for methanol, ethanol, and isopropanol, respectively. The anharmonicity constant was measured to be $84.8 \pm 2.1$ cm$^{-1}$, $80.2 \pm 5.9$ cm$^{-1}$, and $84.4 \pm 6.8$ cm$^{-1}$ for methanol, ethanol, and isopropanol, respectively. The OH vertical dissociation energy was measured to be $499.4 \pm 18.4$ kJ/mol, $518.0 \pm 56.7$ kJ/mol, and $492.7 \pm 59.9$ kJ/mol. The spectroscopically measured values are compared to thermodynamically measured OH bond dissociation energies. The observed differences in previous measurements of the bond dissociation energies compared to the values reported herein can be explained due to the difference between vertical dissociation energies and bond dissociation energies. If the OH overtone stretching mode is excited in methanol to either the 5th or 6th overtone, the bimolecular reaction between methanol and $O_2$ becomes thermodynamically feasible and could contribute to formation of methoxy and HO$_2$ radical under the proper combination of pressure and temperature.

## 1. Introduction

The study and analysis of vibrational overtones in molecules containing an X-H (where X = O, N, C) bond has attracted the attention of both experimental and computational researchers [1–4]. Vibrational overtone transitions have non-zero intensities because of the anharmonicity factor, which arises due to the anharmonicity of bond potentials and non-linearity of dipole functions. The energy absorption primarily occurs in the vibrational overtone levels of X-H stretching vibrations in the near-IR and visible spectral region [2]. These spectra are often well characterized by the local mode (LM) model, which describes the X-H bond as a set of loosely coupled anharmonic oscillators in a highly vibrationally excited molecule [2,5]. Because H is light compared to X (as well as the rest of the molecule), a local mode treatment of the X-H bond allows us to treat the X-H stretch with an approach similar to a diatomic molecule. These transitions have 4–6 orders of magnitude less intensity than electronic transitions. Although these vibrational transitions are weak and less intense than electronic transitions in the same spectral region, these transitions can still be observed for molecules containing X-H bonds. The X-H vibrational stretching overtone spectra provide information about the influence of different local environments between bonds and helps to understand the factors affecting the strength of individual chemical bonds.

The study of O—H vibrational overtone transitions is crucial to

---

understanding some photolytic reactions occurring in the atmosphere [3]. O—H stretching overtone transitions are sharper, narrower, and have well-defined rotational structures compared to most C—H and N—H bonds for similar-sized hydrocarbons [2]. The relatively narrow bandwidth of O—H stretching overtone transitions explains the weaker coupling of O—H bond vibrations to other vibrational modes in alcohols. Several studies have analyzed and reported the intensities of different vibrational overtone transitions of various gas-phase atmospheric species commonly found in the atmosphere [6]. The hydroxyl (OH) radical is one of the most predominant species present in the atmosphere and frequently drives the photochemical oxidation of volatile organic compounds (VOCs) present in the atmosphere to form pollutants such as formaldehyde (HCHO) [7]. It has been observed that overtones can play a substantial role in the photochemistry responsible for OH radical production in the atmosphere [3,8].

Fang et al. [2] successfully studied the conformers of gaseous ethanol by utilizing its gas-phase vibrational overtone spectra. Philips et al. [1] successfully measured the 2nd and 3rd overtone intensities for gas-phase ethanol and isopropanol and up to the 4th overtone in methanol. This work was further extended by Lange et al. [4], who measured the integrated absorption cross-sections of O—H stretching bands for six alcohols, including methanol, ethanol, 1-propanol, 2-propanol, *tert*-butyl alcohol, trifluoroethanol, and two acids, namely acetic and nitric acids. Recently, Wallberg et al. [9] reported the absolute intensities of the fundamental and first overtone transition for various alcohols, including methanol, ethanol, 2-propanol, 1-propanol, and *tert*-butanol.

Measurement of vibrational overtones of alcohols allows for calculation of the vertical dissociation energy of the O—H bond When only the first anharmonicity constant ($\omega_e\,x_e$) is considered, successive vibrational states are given according to Eq. (1).

$$E_{(v)} = \omega_e(\nu+1/2) - \omega_e\chi_e(\nu+1/2)^2 \qquad (1)$$

This expression for energy is the exact same expression as obtained if one solves the Schrödinger equation using a Morse potential [10,11]. Eq. (1) can be used to calculate the vertical dissociation energy using either a Birge-Sponer analysis or by a more recently described method of fitting the overtone energies to a second-order polynomial [12].

In the Birge-Sponer analysis the difference in successive vibrational states is calculated by taking the difference in the successive overtone transitions using Eq. (1), $\Delta E = E_{\nu+1} - E_\nu \omega_e$. This linear expression is plotted verses the appropriate vibrational state to extract vibrational constants. The Birge-Sponer analysis assumes that the sum of successive vibrational energy separations, $\Delta E_\nu$, from the zero-point level to the dissociation limit is the vertical dissociation energy (VDE) or $D_o$. The area under the plot of $\Delta E_\nu$, verses $\nu + 1$ is equal to the sum and therefore $D_o$. Due to the experimental difficulties in measuring higher energy terms (e.g., overtones), typically this plot is linearly extrapolated to higher energy terms. Because the Birge-Sponer analysis with first order anharmonicity generates a linear function, plotting data and extracting a dissociation energy is possible without the use of a computer. Although the linear analysis is straight forward, taking differences between the measured spectroscopic transitions increases the noise in the data used for the analysis. The Birge-Sponer analysis also relies on the accurate assignment of spectroscopic transitions with inaccurate assignments of a transition can have a significant impact on the calculated dissociation energy.

Alternatively, Eq. (1) can be used to find the frequency of the vibrational overtones directly by taking the difference between the energy of the v state and the energy of the v = 0 state, which leads to a quadratic expression for the energy of the spectroscopic transitions. The energy of the vibrational transitions (aka the energy the vibrational states minus the zero point energy) can be fit directly to a second order polynomial. In contrast to use of the linear Birge-Sponer analysis, a fitting of the measured transitions to the quadradic expression neither amplifies the noise in the data, nor is the calculated dissociation energy as sensitive to correct assignment of the spectroscopic transitions, which results in a more accurate determination of the dissociation energy [12].

Historically, BDEs have been measured by use of one of three experimental methods: radical kinetics, gas-phase acidity cycles or photoionization mass spectrometry [13]. Engelking et al. [14] utilized photoelectron spectrometry to investigate the methoxy radical to determine its electron affinity and the vibrational frequencies produced in the detachment process. Additionally, Meot-Ner et al. [15] measured the gas-phase acidity difference between water and methanol using pulsed high-pressure mass spectrometry and defined the acidity of methanol at 300 K. Based on the electron affinity/acidity measurements done by Engelking et al. [14] and Meot-Ner et al. [15], Berkowitz et al. [13] reported the O—H BDE in methanol to be 435.97 ± 3.77 kJ/mol. Recently, Rayne et al. [16] conducted Gaussian calculations and compared their calculated BDE of methanol (436.81 kJ/mole) with that of the recommended experimental value (435.97 ± 3.77 kJ/mol) at 298 K [13,17]. Moreover, Ruscic [18] published Active Thermochemical Tables (ATcT) thermochemistry for BDEs of methanol, methane and ethane and reported the O—H bond dissociation energy in methanol at 298.15 K to be 440.24 kJ/mol. These BDEs values reported for methane, ethane and methanol systems were not measured values but obtained by analyzing and solving a large thermochemical database. These values are claimed to be the most accurate thermochemical values currently available [18]. It is important to note that BDE values represent the difference in enthalpy between the parent molecules and the fully relaxed fragments. Spectroscopic measurements of the vibrational levels assume the Born-Oppenheimer approximation holds, and that we are measuring the vertical dissociation energy difference between the parent molecule and the unrelaxed fragments. Spectroscopic excitation to the vertical dissociation limit in the Born-Oppenheimer approximation leads to direct dissociation of the bond without the possibility of any vibrational relaxation or dissociation to other product channels. Fig. 1 shows the relationship between the vertical dissociation energy, $D_o$, measured spectroscopically and the bond dissociation energy values reported in the ATcT. This difference in energy is the energy of relaxation of the RO fragment from its initial geometry in the alcohol. As is seen in Fig. 1, the vertical bond dissociation value measured by use of either the Birge-Sponer analysis or fit to the second-order polynomial for the O—H bond assumes that the potential energy surface adheres to the Born-Oppenheimer approximation in that the motion of the atomic nuclei and electrons are separated from one another and that the molecular fragments after breaking the O—H bond do not have a chance to relax and move to a lower energy state as is possible when measuring bond dissociation energies using thermodynamic techniques.

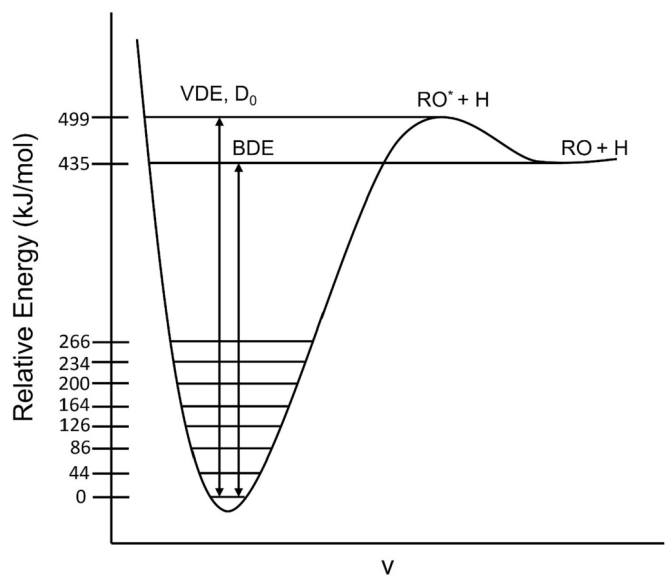

**Fig. 1.** A visualization of a potential energy well showing the difference in energy between the vertical dissociation energy (VDE) and bond dissociation energy (BDE) of a generic alcohol (values shown are for methanol). Pumping any of the alcohols interrogated in this work to the VDE limit results in formation of a vibrationally excited RO* fragment and H. Allowing the RO* fragment to vibrationally relax to RO yields the difference between the VDE and BDE energy.

Absorption spectroscopy has become a widely used approach for the *in-situ* detection of gas-phase species concentration. Quantification of their concentrations using absorption spectroscopy relies on knowledge of their absorption cross-sections [19]. Absorption spectroscopy has several advantages over other detection techniques which makes it an important detection method for gas-phase species. These include *in situ* real-time measurement and high spatial and temporal resolution. Additionally, if absorption cross-sections are known, standards are not necessary to measure concentrations. Different absorption spectroscopy techniques such as cavity ring-down spectroscopy (CRDS) [20,21], cavity-enhanced absorption spectroscopy (CEAS) [22], and incoherent broadband cavity-enhanced absorption spectroscopy (IBBCEAS) [23] have been employed to measure trace gas concentrations and absorption cross-sections over different spectral ranges. These methods are favorable in the quantification of cross-sections due to their longer path length capabilities when compared to other techniques. These longer pathlengths rely on the logic of the Beer-Lambert law to increase sensitivity in measurements.

IBBCEAS, first reported in 2003 by Fiedler et al. [23], has emerged as one highly sensitive methods for detecting gases. Fiedler et al. used an incoherent light source (i.e., a short xenon arc lamp) to measure the concentration of molecular oxygen and gas-phase azulene [23] under atmospheric conditions. Since then, IBBCEAS has been used in the studies for the detection of different gases and their wavelength-dependent cross-sections in the atmosphere [21,24,25]. Using IBB-CEAS, a broadband incoherent light source is used and coupled into a stable optical cavity formed by two highly reflective mirrors. A small fraction of light that leaks from the back end of the optical cavity is then dispersed by a grating and detected with the help of a charged-couple device (CCD) or a photodiode array (PDA).

IBBCEAS offers several advantages in terms of selectivity, sensitivity, compactness, mechanical stability, and time resolved measurements compared to other absorption techniques. Additionally, IBBCEAS has a wide range of practical applications, including (i) simple experimental setup; (ii) does not require any sophisticated and complicated electronic accessories such as fast optical switches, feed-back gates, or loops; (iii) unlike in CEAS, no mode-hop-free scanning is required (iv) unlike diode-lasers which are unable to access short wavelength region, IBBCEAS has the ability to cover a wide range of the electromagnetic spectrum ranging from the short wavelength region of the UV (190 nm) to the infrared region (10 μm); (v) due to the broadband nature of the IBBCEAS technique, many gas-phase species can be detected at the same time. Because of these advantages IBBCEAS has been used to measure the absorption cross-sections for the 5th and 6th overtones transitions for gas-phase methanol, ethanol, and isopropanol.

## 2. Methods

### 2.1. Description of the IBBCEAS instrument setup

Fig. 2 shows the experimental setup. The custom-designed and built IBBCEAS instrument consists of a LED light source, optical filters, a collimating lens, a closed cavity made of highly reflective mirrors, and a grating spectrometer. This study used two IBBCEAS instruments with different path lengths and LED light sources. Measurement of the 6th overtone ($7 \leftarrow 0$) absorption cross-section for methanol, ethanol, and isopropanol was performed using a blue LED light source coupled into a 98.5 cm long cavity, whereas 5th overtone ($6 \leftarrow 0$) measurements were carried out using a 96.5 cm long optical cavity coupled to a green LED light source. Each optical cavity was constructed slightly differently. The 430–490 nm cavity was a 2″ OD PVC tube with nylon mirror mounts. The 483–549 nm cavity was made of a 1″ diameter PFA tube with stainless steel mirror mounts. Both had ¼″ OD input and output ports for gases to be introduced into and out of the cavity. A LED light source was used to produce broadband radiation in these systems. The light was produced by either a blue (LEDEngin) or green LEDs (Thorlabs) centered at 450 and 530 nm, respectively. This light was collimated and introduced into the cavity using a focusing lens. The collimation lenses were f/1 and 2″ diameter (Fresnel) in the 430–490 nm cavity and 1″ diameter in the 483–549 nm cavity. The optical cavity consisted of two 5 cm diameter highly reflective mirrors (reflectivity, R > 99.9 %) (Advanced Thin Films (ATFilms)). The light that leaks through the far end of the cavity was collected and focused onto an optical fiber (1000 μm diameter, 2 m, low-OH, Thorlabs). The absorption cross-sections of these alcohols were measured at a temperature of 25 °C and a pressure of 654 torr. A cage system constructed of carbon-fiber tubes was employed to obtain optical alignment, with structural parts being 3-D printed (laser-sintering or extruded PLA, depending on the function of the part). The carbon-fiber tubes for the smaller triangle cavity (96.5 cm length) were 3/8″ diameter and for the larger cavity (98.5 cm length), 0.5″ diameter carbon-fiber tubes were used. The lengths of the carbon-fiber tubes were about 120 cm for both optical cavities.

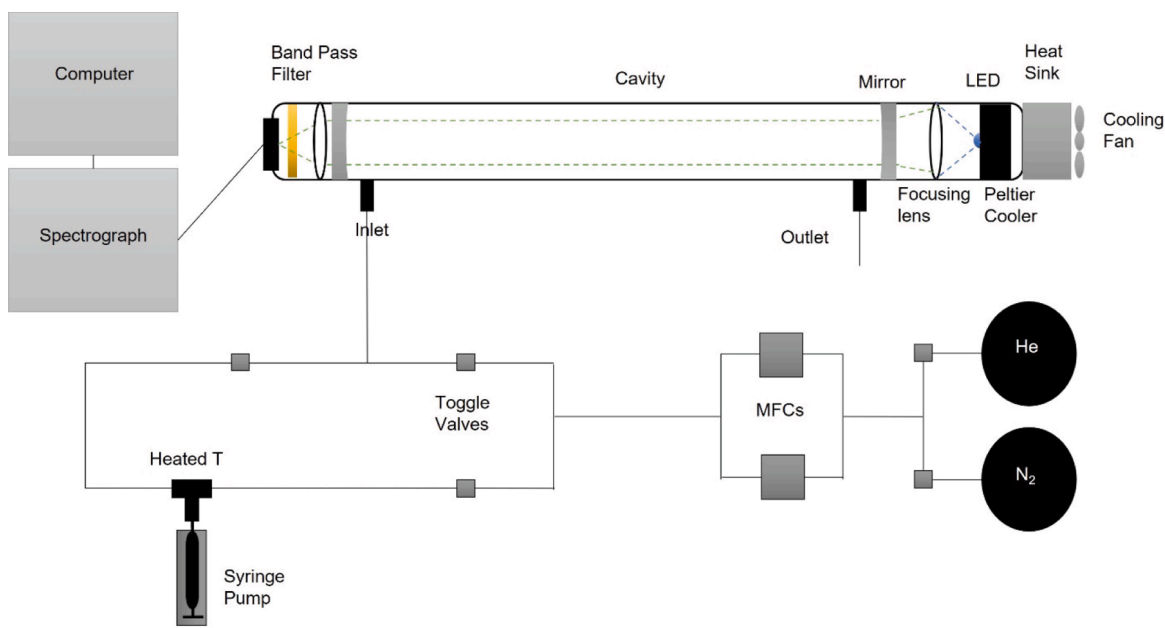

**Fig. 2.** Schematic of the IBBCEAS experimental setup.

Spectra were acquired using toroidal imaging optics using a 303 mm focal length Czerny-Turner spectrometer (Andor Shamrock SR-303iA). Attached to the spectrograph is a charged coupled device cooled to −20 °C (Andor iDus DU440A-bv). The spectrometer was set at the grating groove density of 1200 l/mm (300 nm blaze) to the spectral range of interest. Wavelength calibration was done using a mercury (Hg) pen lamp with a slit width of 50 μm. With these settings, the collected spectra had a resolution of 1 nm.

### 2.2. Workflow of the BBCEAS instrument

The IBBCEAS setup, including LED, spectrometer, and all other components, were allowed to equilibrate for 20 min before data collection. Thermal equilibration allowed for temperature stabilization of the IBBCEAS instrument and minimized fluctuations during spectra collection. Background spectra were acquired at the beginning, middle, and end of each trial to characterize the noise in the CCD detector. This was necessary as the CCD showed non-zero signals even when there was no light impinging on it [26]. A typical trial involved collecting the emission spectra with He (99.99 %, AirGas), N₂ (99.99 %, AirGas), and then either methanol, ethanol, or isopropyl alcohol flowing through the cell. Alcohols were purchased from Sigma-Aldrich and used without further purification. Their stated purities are methanol (HPLC grade, ≥99.9 %), ethanol (anhydrous, gradient grade for liquid chromatography), and isopropanol (HPLC grade, 99.9 %). Alcohols were introduced into the cell by use of a stainless-steel 1/4″ T fitting (Swagelok) packed loosely with glass wool and wrapped with heating tape. A 50 mL gastight syringe (Hamilton 85020) mounted on an automated syringe pump (K_D Scientific KDS100) was used to pump the alcohol of interest into the stainless-steel T at a controlled rate. The temperature of the stainless-steel T was maintained between 115 and 130 °C. This temperature range ensured the vaporization of the alcohols but was below the thermal degradation temperature. Nitrogen gas was used as a carrier gas to sweep the vaporized alcohol from the T-fitting into the cavity. Conditions were chosen to ensure that the alcohol concentration in the gas phase was at least a factor of two less than the saturated vapor pressure of the alcohol.

Spectra were collected using Andor's SOLIS 64-bit software program. A Gaussian smoothing function was applied over a range of 1.5 nm after data collection. A study of the fine rotational structures of the alcohols used was beyond the scope of this paper, and hence, no attempt was made to resolve them. Nitrogen and helium were supplied to the cavity to characterize mirror reflectivity. Mirror reflectivity was calculated as follows:

$$R(\lambda) = 1 - d_0 \left( \frac{\frac{I_{N2}(\lambda)}{I_{He}(\lambda)} \alpha_{Ray}^{N2}(\lambda) - \alpha_{Ray}^{He}(\lambda)}{1 - I_{N2}(\lambda)/I_{He}(\lambda)} \right) \quad (2)$$

where $d_0$ is the cavity length, $I$ is the intensity (spectrum) in nitrogen or helium, and $\alpha$ is the wavelength-dependent Rayleigh scattering.

The change in intensity transmitted through the cavity is related to the extinction due to absorption ($\alpha_{abs}$), calculated by the following mathematical relationship [26]:

$$\alpha_{abs}(\lambda) = \left( \frac{1 - R(\lambda)}{d_0} + \alpha_{Ray}(\lambda) \right) \left( \frac{I_0(\lambda) - I(\lambda)}{I(\lambda)} \right) \quad (3)$$

where $R(\lambda)$ is the wavelength-dependent mirror reflectivity, $d_0$ is the cavity length, $I_0(\lambda)$ is the reference spectrum, $\alpha_{Ray}$ is the wavelength-dependent Rayleigh scattering extinction, and $I(\lambda)$ is the measured spectrum for the absorbing species in the cavity. The wavelength-dependent Rayleigh scattering values for both N₂ and He used were reported by Thalman et al. [25,27].

The absorption cross-sections in cm²/molecule were calculated for different alcohol flow rates by dividing $\alpha_{abs}$ by the corresponding concentration in molecules/cm³ for each flow rate used. The experimental setup is estimated to have the following uncertainties: flow rate (0.1 %), wall interactions (6 %), temperature measurements and fluctuations (5 %), absorbance measurement (10 %), cavity length (1 %), and alcohol concentration (2 %). Random and systematic errors were combined in quadrature to give experimental uncertainties in the experimental values.

## 3. Results & discussion

### 3.1. Absorption Cross-Sections

Figs. 3 to 8 show plots of the absorption cross-sections for the 5th (6 ← 0) and 6th (7 ← 0) overtone transitions in methanol, ethanol, and isopropanol. These measurements provide information about the shape of the energy well for the OH stretch in each alcohol. The wavelength region between 430 and 490 nm was used to measure the absorption cross-sections of the 6th OH overtone transition, whereas the spectral range between 483 and 549 nm was used for the 5th OH overtone absorption cross-section. The individual rotational transitions of the OH overtone are present but unresolved in the observed spectra.

Tables 1 and 2 show the absorption cross-sections for the 5th (6 ← 0) and 6th (7 ← 0) overtones for each alcohol measured in this study. There is a trend observed across the experimental intensities given in Tables 1 and 2. The overtone cross-section decreases with an increase in the length of the carbon chain attached to the OH group. Moreover, as the carbon chain attached to the OH group changes from primary to tertiary, the OH overtone cross-section decreases. This is consistent with the findings of Lange et al. [4], who measured the absorbance cross-sections of the first three overtones for six alcohols, including methanol, ethanol, 1-propanol, 2-propanol, *tert*-butyl alcohol, trifluoroethanol. It was observed that an increase in the length of the carbon chain corresponds to a decrease in the intensity of the overtone transition.

Additionally, they reported that the change in the intensities of the fundamental transitions for several different compounds was dependent on the electronegativity of the substituent attached to the OH group in the molecules they studied. Although the decrease in the OH fundamental intensities is directly related to an increase in the electronegativity of the group attached to the alcohol, the analogous trend is much less apparent for subsequent higher absorption overtone intensities. The absorption cross-section intensities for the 5th and 6th OH overtones of alcohols measured in this study were found to be shifted towards a longer wavelength with an increase in the alkyl groups attached to the OH group. Alkane chains are recognized as the least strongly polarizing group that still exert some electron-withdrawing capability towards the OH group in which it is attached. As such, there is a measurable decrease in the intensity of the overtone transition with a lengthening of the alkane chain attached to the OH group.

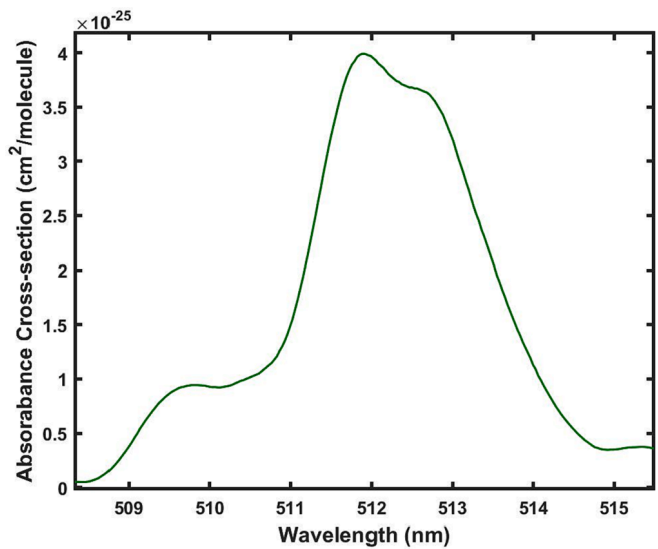

**Fig. 3.** Fifth overtone cross-section of methanol determined between 509 and 515 nm.

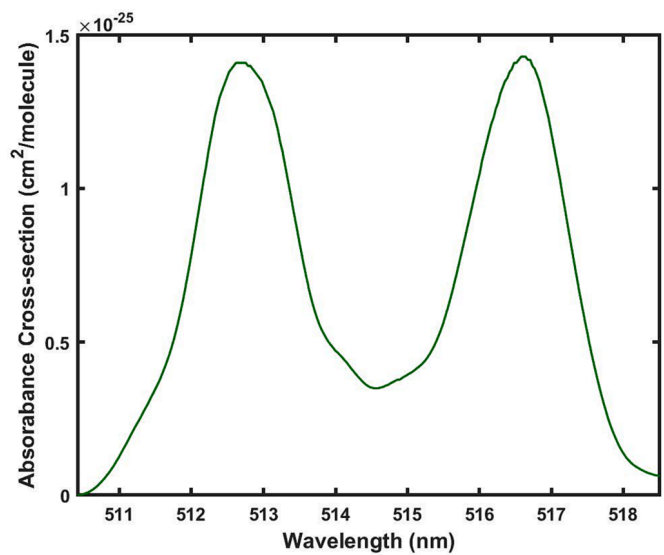

**Fig. 4.** Fifth overtone cross-sections of ethanol determined between 511 and 518 nm.

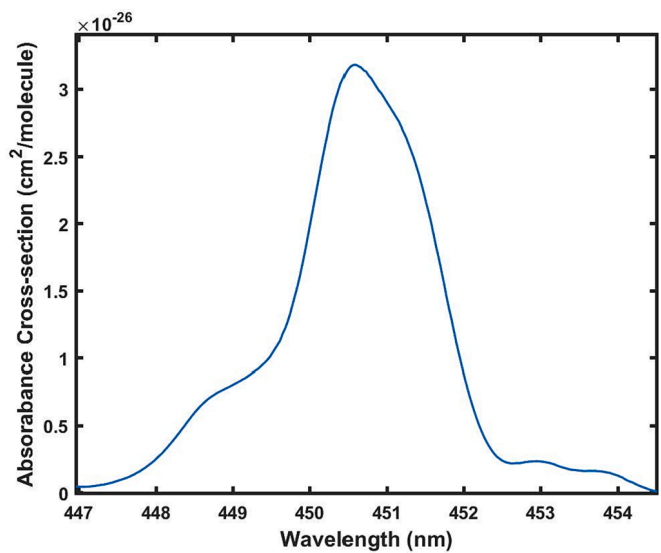

**Fig. 6.** Sixth overtone cross-sections of methanol between 447 and 454 nm.

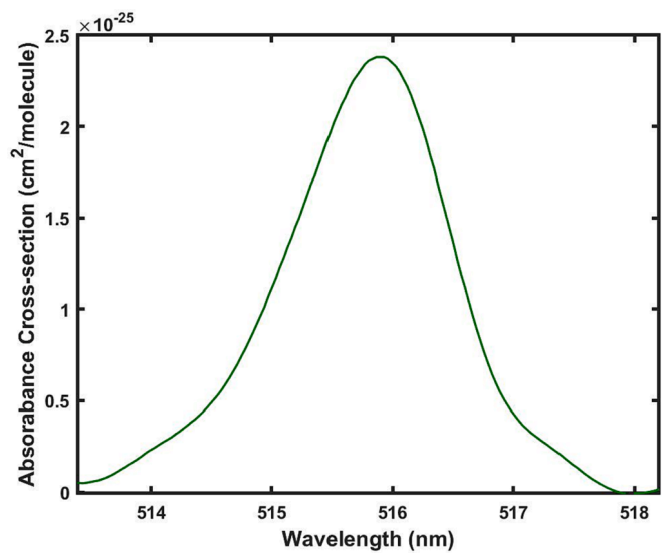

**Fig. 5.** Fifth overtone cross-sections of isopropyl alcohol between 514 and 518 nm.

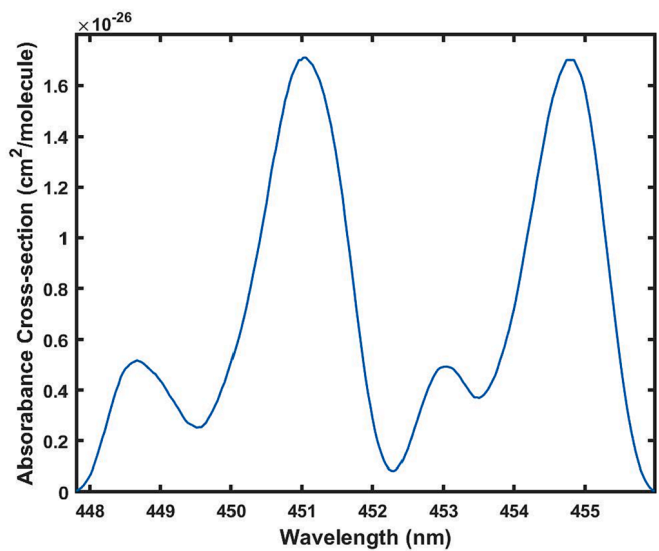

**Fig. 7.** Sixth overtone cross-sections of ethanol between 448 and 456 nm.

Tables 1 and 2 show the values for the measured absorption cross-sections for 5th and 6th overtone of methanol, ethanol, and isopropanol. The wavelength values reported included the absorption cross-section peaks of each alcohol studied.

Based on the formulas given by Phillips et al. [1], band locations of several overtone transitions (up to 9th overtone) were calculated. This reveals information about the center wavelength corresponding to each overtone transition for methanol, ethanol, and isopropanol. The experimental values from this work were compared with the predicted overtone transitions and good agreement has been observed between the values. Notably, double band profiles were observed for ethanol overtone transitions, as shown in Figs. 4 and 7. This can be attributed to the mixture of different conformers of ethanol showing more than one band for 5th and 6th overtone [2]. It has previously been found that the higher energy band is from the trans conformation of ethanol, while the lower energy band is due to the gauche conformation of ethanol. When ethanol is in the trans conformation, the OH stretch is parallel to the molecular axis. In contrast, when ethanol is in the gauche conformation, the OH stretch is perpendicular to the molecular axis. The difference in vibrational energy between the gauche and trans conformation accounts for the two bands observed in the ethanol overtone cross-section. Table 3 shows the predicted and experimentally determined center wavelength for the 5th and 6th overtone transitions for methanol, ethanol, and isopropanol.

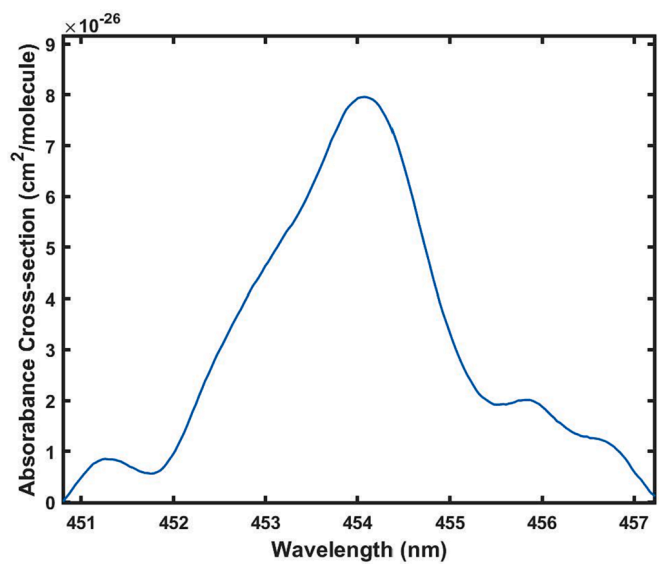

**Fig. 8.** Sixth overtone cross-sections of isopropanol between 451 and 457 nm.

**Table 1**

Fifth overtone cross-sections and band strengths of methanol, ethanol, and iso-propyl determined between 510 and 517.5 nm. The error in these values is 13 %.

| Wavelength (nm) | Measured Methanol Cross-section ($10^{-25}$ cm²/molecule) | Measured Ethanol Cross-section ($10^{-25}$ cm²/molecule) | Measured isopropyl alcohol Cross-section ($10^{-25}$ cm²/molecule) |
|---|---|---|---|
| 510 | 0.93 (0.12) | – | – |
| 511.5 | 2.89 (0.38) | 0.43 (0.06) | – |
| 513 | 2.96 (0.38) | 1.16 (0.15) | – |
| 514.5 | 0.68 (0.09) | 0.41 (0.05) | 0.61 (0.08) |
| 516 | – | 1.01 (0.13) | 1.88 (0.24) |
| 517.5 | – | 0.61 (0.08) | 0.21 (0.03) |
| Band Strength | 7.46 (0.97) | 3.62 (0.47) | 2.70 (0.35) |

**Table 2**

Sixth overtone cross-sections and band strengths of methanol, ethanol, and isopropyl alcohol were determined between 448 and 455.5 nm. The error in these values is 13 %.

| Wavelength (nm) | Measured Methanol Cross-section ($10^{-25}$ cm²/molecule) | Measured Ethanol Cross-section ($10^{-25}$ cm²/molecule) | Measured isopropyl alcohol Cross-section ($10^{-25}$ cm²/molecule) |
|---|---|---|---|
| 448 | 0.31 (0.04) | – | – |
| 449.5 | 1.29 (0.17) | 0.41 (0.05) | – |
| 451 | 2.70 (0.35) | 1.36 (0.18) | 0.34 (0.04) |
| 452.5 | 0.46 (0.06) | 0.33 (0.04) | 2.90 (0.38) |
| 454 | – | 0.88 (0.11) | 6.83 (0.89) |
| 455.5 | – | 0.69 (0.09) | 2.42 (0.31) |
| Band Strength | 4.76 (0.62) | 3.67 (0.48) | 12.49 (1.62) |

### 3.2. Calculation of vertical dissociation energy and constants

Fig. 9 shows the Birge-Sponer plot for methanol, ethanol, and isopropanol, along with the 95 % confidence limit intervals. This plot is used to fit the differences in successive OH overtone transition energies for an anharmonic oscillator. Mathematically, the fitting of the OH overtone transition energies differences for anharmonic oscillators is given by the one-dimensional Birge-Sponer equation obtained by taking the differences between the $0 \rightarrow v$ and the $0 \rightarrow v + 1$ transitions using Eq. (1) as follows [2]:

$$\Delta E_v = \omega_e - 2\omega_e \chi_e(v + 1) \tag{4}$$

where $\Delta E_v$ represents the energy difference between the $v$ and the $v + 1$ quantum level, parameters $\omega_e$ and $\omega_e x_e$ represent the fundamental frequency and anharmonicity of the oscillator, respectively. This type of analysis for determining bond dissociation energies is known to over-estimate this value. This is because this analysis assumes a linear extrapolation towards higher energy terms, which ignores higher order anharmonicity. Indeed, Phillips et al. speculated that there may be substantial curvature in a Birge-Sponer plot due to existence of combination bands, as observed by Fang et. al. in their study of vibrational overtones of gaseous alcohols. Inclusion of the high order overtones measured in this work into the Birge-Sponer plot for methanol, ethanol and isopropyl alcohol shows this plot to be linear and not suffering from the issues previously raised through the v = 6 transition.

Integration of the area under the curve of the Birge-Sponer plot yields the vertical dissociation energy of the O—H bond for each alcohol. The O—H vertical dissociation energy in methanol, ethanol and iso-propanol were calculated using this plot. Additionally, the slope of these plots yields the first order anharmonicity constant of the OH stretch in each alcohol and the intercept gives the fundamental vibrational frequency. These values are reported in Table 4-6. These findings agree with thoses of Fang et al. [2], who reported that the anharmonicities for all alcohols are essentially identical (86 cm$^{-1}$).

Note the great degree of uncertainty in these linear plots show by the 95 % confidence limit intervals. This large uncertainty results from plotting differences in transitions, which amplifies the uncertainty of the values used to extract the spectroscopic constants.

Birge-Sponer plots have traditionally been used to determine molecular constants and energies for vibrational transitions [28]. With the advent of better fitting software, the necessity of linearizing a function for ease of use has diminished. It has been found that fitting overtone data to a second-order polynomial yields more accurate results, as seen in Fig. 10 [29]. Some errors created by taking difference in spectroscopic transitions can be avoided by plotting a second-order polynomial as the relationship between the energy of transitions. This method fits the transition frequencies to a second-order polynomial to yield molecular constants and energy terms. This equation is

$$\nu = a\left(v + \frac{1}{2}\right)^2 + b\left(v + \frac{1}{2}\right) + c \tag{5}$$

where a = $-\omega_e x_e$, b = $\omega_e$, and c = $-\left(\frac{\omega_e}{2} - \frac{\omega_e x_e}{4}\right)$. Eq. (5) is found by found by applying Eq. (1) to the overtone transitions, $0 \rightarrow v$. $\omega_e x_e$ is the

**Table 3**

Comparison of center wavelength predicted from formulas given in Phillips et al. versus the experimental measurements presented here for the 5th and 6th OH overtone transitions of methanol, ethanol, and isopropanol. Two values are reported for ethanol overtone center wavelength due to the double band profile in overtone transitions. The error in these values is 5%.

| Transition | Methanol OH overtone center wavelength (nm) | | Ethanol OH overtone center wavelength (nm) | | Isopropanol OH overtone center wavelength (nm) | |
|---|---|---|---|---|---|---|
| | Predicted | Experimental | Predicted | Experimental | Predicted | Experimental |
| 6 ← 0 | 512.2 | 511.85 | 515.0 | 512.49 / 516.80 | 518.2 | 516.08 |
| 7 ← 0 | 450.9 | 450.5 | 453.7 | 450.88 / 454.84 | 456.9 | 454.4 |

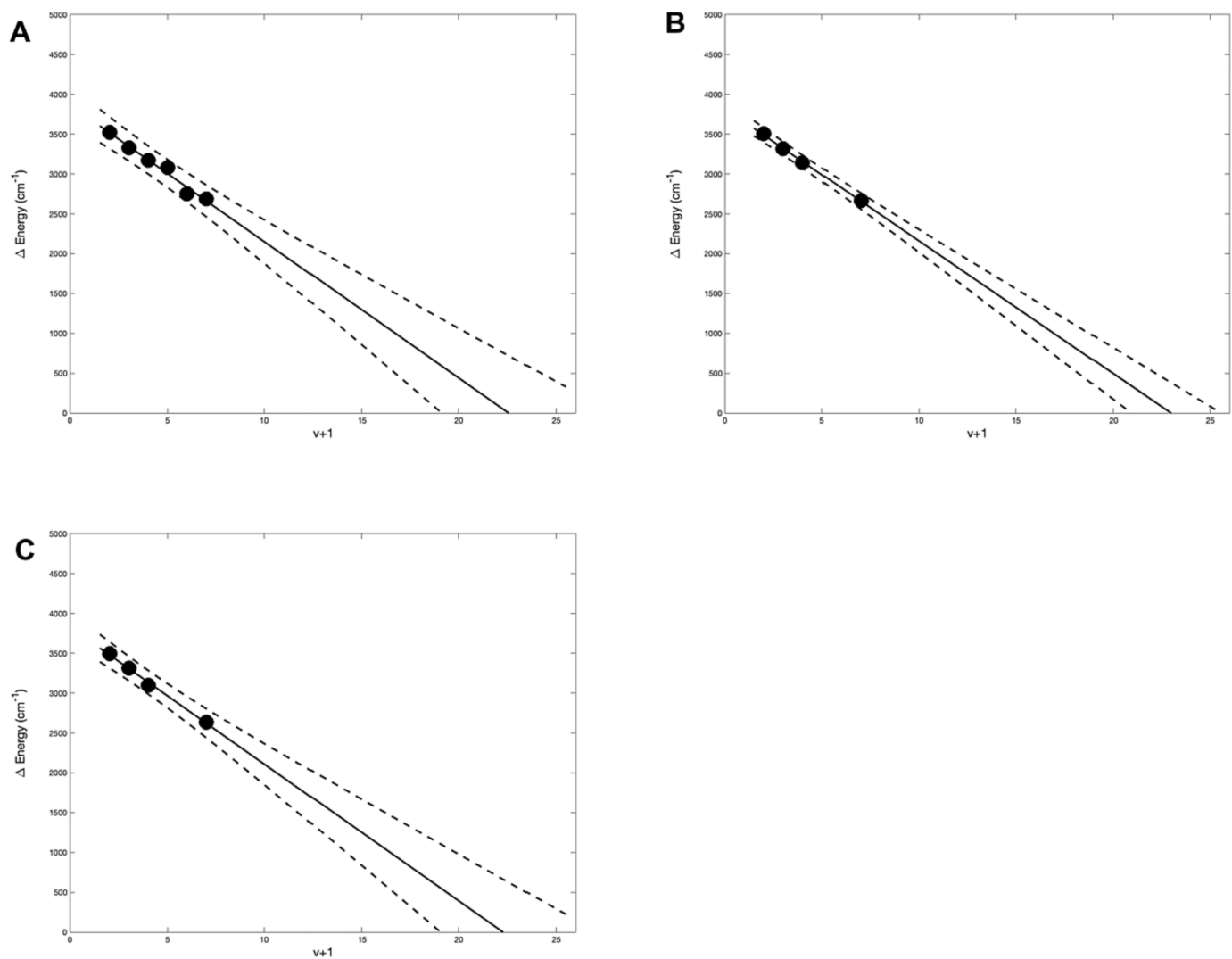

**Fig. 9.** Birge-Sponer plot for the OH stretch of (A) methanol (y = -171x + 3860), (B) ethanol (y = -171.4x + 3822), and (C) isopropanol (y = -166.4x + 3823). The dashed lines in the fit show the 95 % confidence intervals. The first four overtones were calculated by Phillips et al. The 5th and 6th overtones were experimentally measured in this work.

**Table 4**
Molecular constants and dissociation energies calculated using both the Birge-Sponer analysis and second-order polynomial fit for methanol. Uncertainty in parentheses.

| Constant | Polynomial Fit | Birge-Sponer |
|---|---|---|
| $\omega_e$ (cm$^{-1}$) | 3848(19) | 3860(185.5) |
| $\omega_e x_e$ (cm$^{-1}$) | 84.8(2.1) | 85.5(19.3) |
| VDE (kJ/mol) | 499.4(18.4) | 498.3(175.8) |

**Table 5**
Showing the molecular constants and dissociation energies calculated using both the Birge-Sponer analysis and second-order polynomial fit for ethanol. Uncertainty in parentheses.

| Constant | Polynomial Fit | Birge-Sponer |
|---|---|---|
| $\omega_e$ (cm$^{-1}$) | 3807(55) | 3822(156) |
| $\omega_e x_e$ (cm$^{-1}$) | 80.2(5.9) | 857.(17.7) |
| VDE (kJ/mol) | 518.0(56) | 487.1(152.3) |

**Table 6**
Showing the molecular constants and dissociation energies calculated using both the Birge-Sponer analysis and second-order polynomial fit for isopropyl alcohol. Uncertainty in parentheses.

| Constant | Polynomial Fit | Birge-Sponer |
|---|---|---|
| $\omega_e$ (cm$^{-1}$) | 3813(63) | 3823(86.5) |
| $\omega_e x_e$ (cm$^{-1}$) | 84.4(6.8) | 83.2(9.8) |
| VDE (kJ/mol) | 492.7(59.9) | 502.5(86.6) |

fundamental vibrational frequencies, and anharmonicity constants calculated using this method are also reported in Tables 4-6. The maximum vibrational state can be found from Eq. (5) to be

$$v_{max} = -\frac{a + b}{2a} \tag{6}$$

Eq. (6) works under the assumption, discussed by Lessinger et al, that the vibrational quantum number is a continuous function [12].

As a note, because the quadradic expression for the energy of vibrational states given by Eq. (1) results from a Morse potential and the energy expression used to fit the overtones (Eq. (5)) comes directly from Eq. (1), the values obtained from the fits in Fig. 10 are identical to those obtained from the Morse expression, $D_e = \frac{\omega_e^2}{4\omega_e x_e}$ and $D_o = D_e - (\frac{\omega_e}{2} - \frac{\omega_e x_e}{4})$.

anharmonicity constant in the ground electronic state, and $\omega_e$ is the fundamental frequency in the ground electronic state and c represents the zero point energy of the state. Vertical dissociation energies,

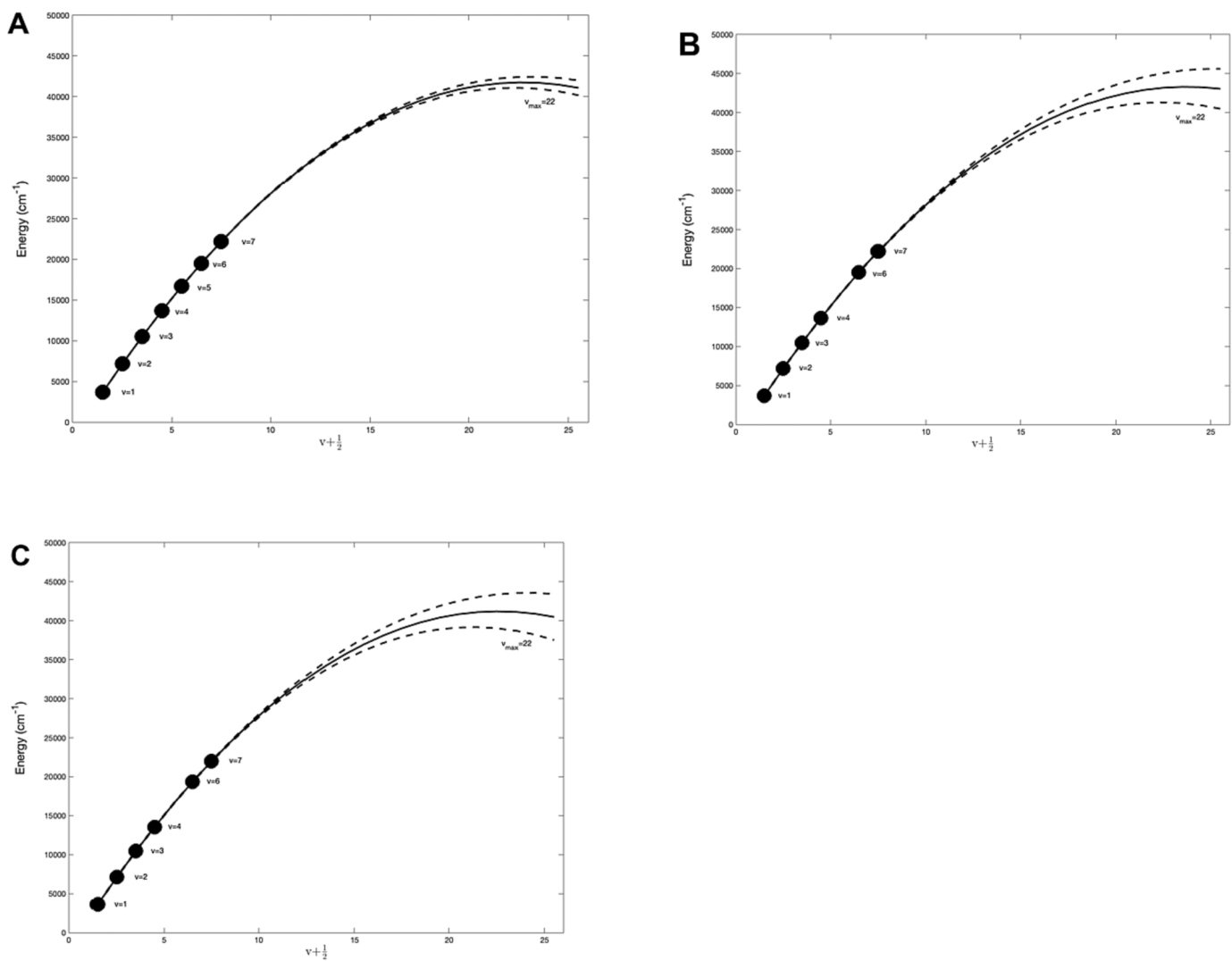

**Fig. 10.** Second-order polynomial fit to the measured overtone data and used to calculate $\nu_{max}$, the anharmonicity constant, and fundamental constant for (A) methanol ($y = -84.2x^2 + 3848x-1895$), (B) ethanol ($y = -80.2x^2 + 3807x-1853$), and (C) isopropyl alcohol ($y = 84.4x^2 + 3813x-1870$). The dashed lines in each of the plots shows the 95 % confidence intervals of the fits.

### 3.3. Comparison of molecular constants and energy terms using Birge-Sponer and Second-order polynomial fit

As explained, both the Birge-Sponer analysis used here and the second-order polynomial fit to the data both only include first order anharmonicity and if there is no uncertainty in the data, both would give the exact same results; however, the Birge-Sponer analysis is performed by taking the differences between successive states, rather than using the frequencies of the spectroscopic transitions directly, which increases the noise in the data used for the analysis. The second order fit uses the spectroscopic transitions directly and is subject to less noise in the analysis. As a result, the calculation of VDE using the Birge-Sponer analysis has a greater degree of uncertainty than does VDE calculated using the second-order polynomial method, which is clearly seen in the 95 % confidence limit intervals in Figs. 9 and 10. Both methods require one to determine the vibrational state at the convergence limit. Because the Birge-Sponer analysis does this using a linear extrapolation it is very sensitive to the correct assignment of the transitions and being off by even one state in assignment can result in a large error of VDE. On the other hand, the determination of the convergence limit using the second-order polynomial method is nearly the same when spectral assignments are assumed to be one off on either side of the correct assignments. Tables 4-6 show the calculated values for the fundamental vibrational frequency ($\omega_e$), anharmonicity constant ($\omega_e x_e$), and vertical dissociation energy (VDE) for methanol, ethanol, and isopropyl alcohol. These values are calculated from the fit line functions generated by each graph which are extrapolated to calculate the convergence limit and its vibrational state. The fit for the polynomial yields a more accurate value than that of the Birge-Sponer plot. These differences can be seen in the tables below. As we are able to eliminate some uncertainty created by the assumption of linearity in the Birge-Sponer plots, we have chosen to use the second-order polynomial fit for our analysis.

### 3.4. $CH_3OH^* + O_2 \rightarrow$ products

Recently research has been done to understand the importance of overtone excitations of OH vibrations in atmospherically important molecules. There have been several studies in which molecules are vibrationally excited to higher overtones to measure the dissociation energy of various functional groups in a molecule [30]. In addition, the possibility of chemical reactions occurring because of vibrational overtone excitation has also been explored [31]. These overtone excitations can lead to the dissociation of excited molecules and may result in the formation of new molecules or radicals in the atmosphere. Donaldson et al. [32] showed that the OH vibrational overtone excitation of hydrogen peroxide ($H_2O_2$) to higher overtones ($\nu_{OH} \geq 4$) allows the $H_2O_2 + O_2$ reaction to overcome the activation energy barrier to form $HO_2 + HO_2$. This study was important in understanding the potential role of vibrationally excited HOOH in recycling $HO_x$ ($HO_x = OH + HO_2$) when reacted with $O_2$. Although the contribution of overtone-driven reaction of $HO_x$ production might not be large, the contribution of this mechanism becomes much more significant at high solar zenith angles, increasing the $HO_x$ production to about 20–30 % by this mechanism [32].

It is hypothesized that pumping methanol to the 4th, 5th, or 6th overtone may enable the methanol + oxygen reaction to overcome the activation energy barrier to form radical products.

Fig. 11 shows a slice of the potential energy surface for the reaction of $CH_3OH + O_2$. $O_2$ can react with vibrationally excited $CH_3OH$ (e.g., $CH_3OH^*$) to extract either a methyl hydrogen or the alcohol hydrogen. Previous work reports that these activation barriers to range between 149 and 188 kJ/mol [33]. Superimposed on the figure are the $v = 1–7$ energy levels for OH stretch in $CH_3OH$. As can be seen, exciting $CH_3OH$ via pumping to either the $v = 6$ or $v = 7$ OH vibrational level introduces enough energy that the reaction can overcome both the activation barrier and endothermicity of the reaction products. Upon excitation of the OH stretch, several outcomes are possible. Methanol can directly react with $O_2$ prior to intramolecular vibrational energy redistribution (IVR) if the methanol + $O_2$ collision rate is faster than IVR, it can be collisionally deactivated, it can undergo IVR decreasing the rate of reaction. The various possible outcomes depend on the relative time scales of the different processes.

Rizzo, Perry, and co-workers have studied intramolecular energy transfer in highly excited methanol and determined IVR times for the $v = 5–8$ OH stretch levels of 130 fs, 3.2 ps, 240 fs, and 200–300 fs, respectively [34,35]. While excitation to $v = 5$, 7 and 8 all relax on the order of 100–300 fs, relaxation from $v = 6$ is an order of magnitude longer. Excitation to this state provides the best opportunity for the enhancement of the methanol and molecular oxygen reaction. It is interesting to see that the lifetime of the energy that stays in the 6th level (i.e. 5th overtone) is 3.2 ps and that the lifetime decreases to 240 fs when pumped to the 6th overtone, and that the lifetime gets shorter when pumped to higher and higher energy levels.

Collision frequencies as a function of pressure for $O_2/CH_3OH$ collisions were calculated for comparison to these IVR Times; thus, the pressure of $O_2$ needed to match the collision rate with the rate of IVR for pumping to both the 5th and 6th overtone was computed. A pressure of 50.52 atm yields a collision rate of 3.2 ps which matches the lifetime of $CH_3OH$ when excited to the 5th overtone, and a pressure of 673.60 atm is required for $O_2$ to collide with methanol on the order of 240 fs (the IVR lifetime of the 6th overtone) at 298 K.

Rate constants for the $CH_3OH + O_2$ reaction have been studied previously using transition state theory at T = 1000 K [36]. Abou-Rachid et al. [36] calculated rate constants for the extraction of either a methyl hydrogen or alcohol hydrogen from methanol by $O_2$. They reported rate constants of 194 and 70 $cm^3$ $mol^{-1}$ $s^{-1}$ for reaction with the methyl hydrogen and the alcohol hydrogen, respectively. While these reactions would occur at much longer times than the IVR times measured by Rizzo and Parry [34], it is important to note that even at T = 1000 K the population of the excited methanol molecules in the $v = 5–8$ states is extremely small. The fraction of methanol molecules in the $v = 5–8$ states is $3.5 \times 10^{-12}$, $6.7 \times 10^{-14}$, $1.4 \times 10^{-15}$ and $3.8 \times 10^{-17}$, respectively. Direct excitation of these modes would significantly increase the rate constants of the reaction between methanol and oxygen, especially if the reaction occurs prior to IVR. Even after all the energy from the excitation has been redistributed in the molecule, the reaction rate will be much faster than those calculated at 1000 K. Equipartitioning of 5, 6 and 7 quanta of OH stretch using the energy the states measured here, correspond to a vibrational temperature of 3125, 3480, and 3820 K, respectively. Additional efforts are currently underway to investigate the effect of overtone excitation in these molecules on reaction rates, both before and after IVR has set in.

## 4. Conclusions

The absorption cross-sections for the 5th and 6th OH stretching overtones for methanol, ethanol, and isopropanol in the gas-phase are reported. The measured center wavelengths of 511.85 nm (methanol), 512.49 and 516.80 (ethanol), and 516.08 nm (isopropanol) for the 5th OH overtone are in good agreement with previously reported calculated values. The 6th OH overtone values for methanol (450.5 nm), ethanol (450.88, 454.84), and isopropanol (454.4 nm) are also in good agreement with computational predictions. The measurements of OH overtone transitions clearly demonstrate an order of magnitude drop in the measured absorption cross-section measurements for the 5th and 6th overtone transitions, consistent with what was predicted by Phillips et al.[1] Despite being the least strong polarizing group, alkanes still possess some electron withdrawing tendencies towards the OH group attached to them and as such are responsible for a measurable decrease in the intensity of the overtone transition. Additionally, it was seen that an increase in the carbon chain length of alcohols shifts the center wavelength of OH overtone absorption cross-sections to a slightly higher wavelength. The measured vibrational OH overtones in methanol coupled with the calculated reaction thermodynamics for the biomolecular reaction of vibrationally excited methanol with $O_2$ shows the reaction to be energetically favorable and under the proper combination of temperature and pressure could contribute to the formation of methoxy and $HO_2$ radicals.

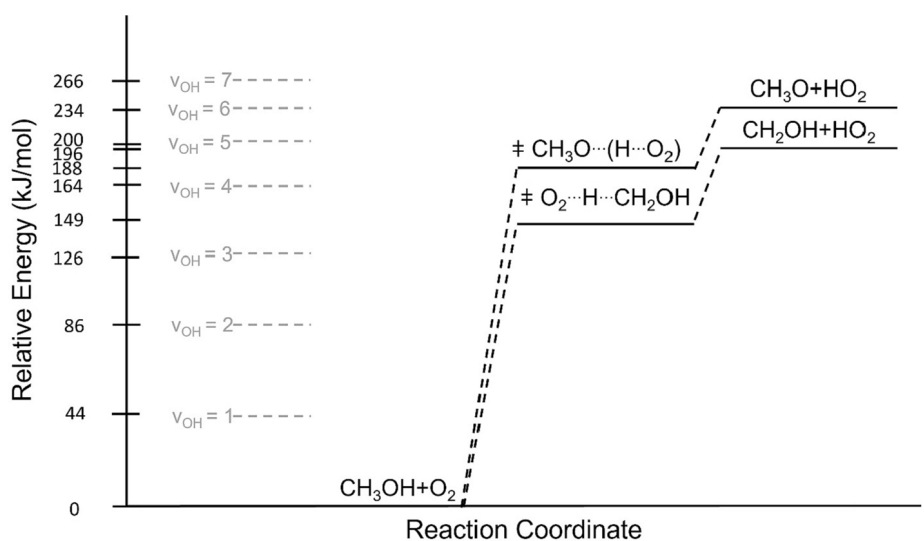

**Fig. 11.** Potential energy surface for the reaction of $CH_3OH^* + O_2$, along with the vibrational energy of the various OH stretch states. The energy levels of the overtones as reported by Phillips et al. [1].

## 5. Author Statement

Callum Flowerday aided in the collection of data, analysis of the data and writing of the manuscript. Nitish Bhardwaj aided in the collection of data, analysis of data and writing of the manuscript. Ryan Thalman aided in the experimental design, and construction of the instruments and in data analysis. Matthew Asplund and Eric Sevy aided in data analysis and writing of the manuscript. Jaron Hansen aided in the experimental design, construction, data analysis and writing of the manuscript.

## Funding

This material is based upon work supported by the National Science Foundation under Grant No. 2114655.

## Declaration of Competing Interest

The authors declare that they have no known competing financial interests or personal relationships that could have appeared to influence the work reported in this paper.

## Data availability

Data will be made available on request.